\documentclass[
    reprint,
    preprintnumbers,
    superscriptaddress,
    nofootinbib,
     amsmath,amssymb,
     aps,
     prd,
    floatfix,
    ]{revtex4-2}
    
    \usepackage{graphicx}
    \usepackage[usenames,dvipsnames]{xcolor} % colour
    \usepackage{times, mathptmx} % Times font
    \usepackage[utf8]{inputenc}
    \usepackage{color}
    \usepackage{hyperref}
    \allowdisplaybreaks[1]
    \graphicspath{{figures/}}

    \newcommand{\mpl}{M_\mathrm{Pl}}
    \newcommand{\rext}{r_\mathrm{ext}}
    \newcommand{\rew}{r_\mathrm{rew}}
    \newcommand{\iu}{\mathrm{i}\mkern1mu}

    \newcommand{\qmul}{Geometry, Analysis and Gravitation, School of Mathematical Sciences, Queen Mary University of London,
    Mile End Road, London E1 4NS, United Kingdom}
    \newcommand{\oxford}{Astrophysics, University of Oxford, Denys Wilkinson Building, Keble Road, Oxford OX1 3RH, United Kingdom}
    \newcommand{\kcl}{King's  College  London,  Strand,  London  WC2R  2LS,  United Kingdom}
    
\begin{document}
        
    \title{Where is the ringdown? Reconstructing quasinormal modes from dispersive waves}
    
    \author{Josu C. Aurrekoetxea}
    \email{josu.aurrekoetxea@physics.ox.ac.uk}
    \affiliation{\oxford}
    
    \author{Pedro G. Ferreira}
    \affiliation{\oxford}
    
    \author{Katy Clough}
    \affiliation{\qmul}
    
    \author{Eugene A. Lim}
    \affiliation{\kcl}
    
    \author{Oliver J. Tattersall}
    \affiliation{\oxford}
    
    % \date{Received \today; published -- 00, 0000}
    
    \begin{abstract}
    
    We study the generation and propagation of gravitational waves in scalar-tensor gravity using numerical relativity simulations of scalar field collapses beyond spherical symmetry. This allows us to compare the tensor and additional massive scalar waves that are excited. As shown in previous work in spherical symmetry,  massive propagating scalar waves decay faster than $1/r$ and disperse, resulting in an inverse chirp. These effects obscure the ringdown in any extracted signal by mixing it with the transient responses of the collapse during propagation. In this paper we present a simple method to \textit{rewind} the extracted signals to horizon formation, which allows us to clearly identify the ringdown phase and extract the amplitudes of the scalar quasinormal modes, quantifying their excitation in strong gravity events and verifying the frequencies to perturbative calculations.  The effects studied are relevant to any theories in which the propagating waves have a dispersion relation, including the tensor case.
    
    \end{abstract}
    
    \maketitle
    
    %%%%%%%%%%%%%%%%%%%%%%%%%%%%%%%%%%%%%%%%%%%%%%%%%%%%%%%%%%%%%%
    
    \section{Introduction}

    The advanced LIGO and Virgo network have given us, for the first time, a window onto the inspiral and merger of black hole binaries and neutron star binaries \cite{LIGOScientific:2016aoc,LIGOScientific:2017vwq}. The hope is that these instruments will allow us to find and characterize other, more exotic, 
    events that generate gravitational waves -- opening a window onto new physics that comes to the fore in strong gravity \cite{Berti:2015itd}.
    
    Of particular interest is the possibility that there are new fields that interact with gravity and, to some extent, modify it. These new fields may lead to long range forces, which complement the gravitational force, or they may propagate, 
    adding an additional channel for energy loss
    and potentially altering the waveforms that emerge from gravitational events. Scalar-tensor theories \cite{Fierz:1956zz,Jordan:1959eg,Wagoner:1970vr,Bergmann:1968ve,Nordtvedt:1970uv,fujii_maeda_2003,Clifton:2011jh,Damour:1998jk}, with one or more scalar fields, have a well-posed Cauchy problem  \cite{Damour:1992we,Damour:1996ke,Salgado:2005hx,Salgado:2008xh,Horbatsch:2015bua} that permits a numerical evolution \cite{Berti:2013gfa}
    \footnote{Other beyond-GR theories with well posed initial value formulations that have been studied numerically include EdGB \cite{Elley:2022ept, Shiralilou:2021mfl, Silva:2020omi, Witek:2018dmd, East:2020hgw, Carson:2020ter, Ripley:2019aqj}, dynamical Chern Simons \cite{Okounkova:2017yby, Okounkova:2020rqw,Okounkova:2019zjf, Okounkova:2019dfo}, $k$-essence \cite{Bezares:2020wkn,terHaar:2020xxb,Bezares:2021yek,Bezares:2021dma} and cubic Horndeski \cite{Figueras:2021abd, Figueras:2020dzx}.},
    and are of particular interest as they arise as effective field theories in a number of different contexts. The workhorse of such theories is Jordan-Brans-Dicke theory with a potential \cite{Brans:1961sx}, but it can be generalized to the Horndeski theories \cite{Horndeski:1974wa,Deffayet:2009wt} and beyond.
    
    Although massless scalar-tensor theories are severely constrained both by solar system experiments \cite{Will:2001mx,Bertotti:2003rm} and binary pulsar observations \cite{Freire:2012mg,Antoniadis:2013pzd}, their massive counterparts remain widely unexplored \cite{Alsing:2011er,Berti:2012bp}. For scalar-tensor theories satisfying the GW170817 constraint on the speed of gravitational waves \cite{Baker:2017hug,Creminelli:2017sry,Ezquiaga:2017ekz} and in which the scalar field plays a significant cosmological role (i.e. in which the energy density of the scalar field is comparable with other dominant constituents of the Universe), black holes will look very much like those in general relativity \cite{Tattersall:2018map} -- they won't have ``hair". The standard lore is, then, that it will be impossible to detect any evidence of new physics, through signatures of the scalar field.

    However, even if the endpoint of gravitational collapse is a black hole with no hair, fluctuations in the scalar field are still possible \cite{Barausse:2008xv,Tattersall:2018nve}. This is the case if the starting point that leads to black hole formation had some non-trivial profile in the scalar field -- there can then be an imprint in the transient behaviour towards the final black hole. In the case of a binary, a scalar environment could affect the inspiral, merger and the ringdown stages of the event. During the ringdown phase, which is primarily characterized in terms of a superposition of damped exponentials with
    complex frequencies known as quasinormal modes (QNMs) \cite{Kokkotas:1999bd,Nollert:1999ji,Berti:2009kk}, there will be a set of additional scalar modes with frequencies that are determined by properties of the scalar field. These can be found using the standard methods for calculating quasinormal modes. But this phenomena -- of extra propagating waves in the gravitational wave signal -- will be true more generally, beyond the quasinormal mode components. Indeed, we will distinguish between the quasinormal \emph{modes} (i.e. perturbations of the metric at source) and the subsequent propagating \emph{waves} in this paper which are sourced by these modes. In the massless tensor sector, this distinction is academic, since they are non-dispersive -- the chronology of detection is exactly the same as the chronology of emission. As we will see, in the case of massive modes, the dispersive nature mixes up this chronology and hence the distinction becomes useful.

    \begin{figure*}[t]
        \includegraphics[width=\linewidth]{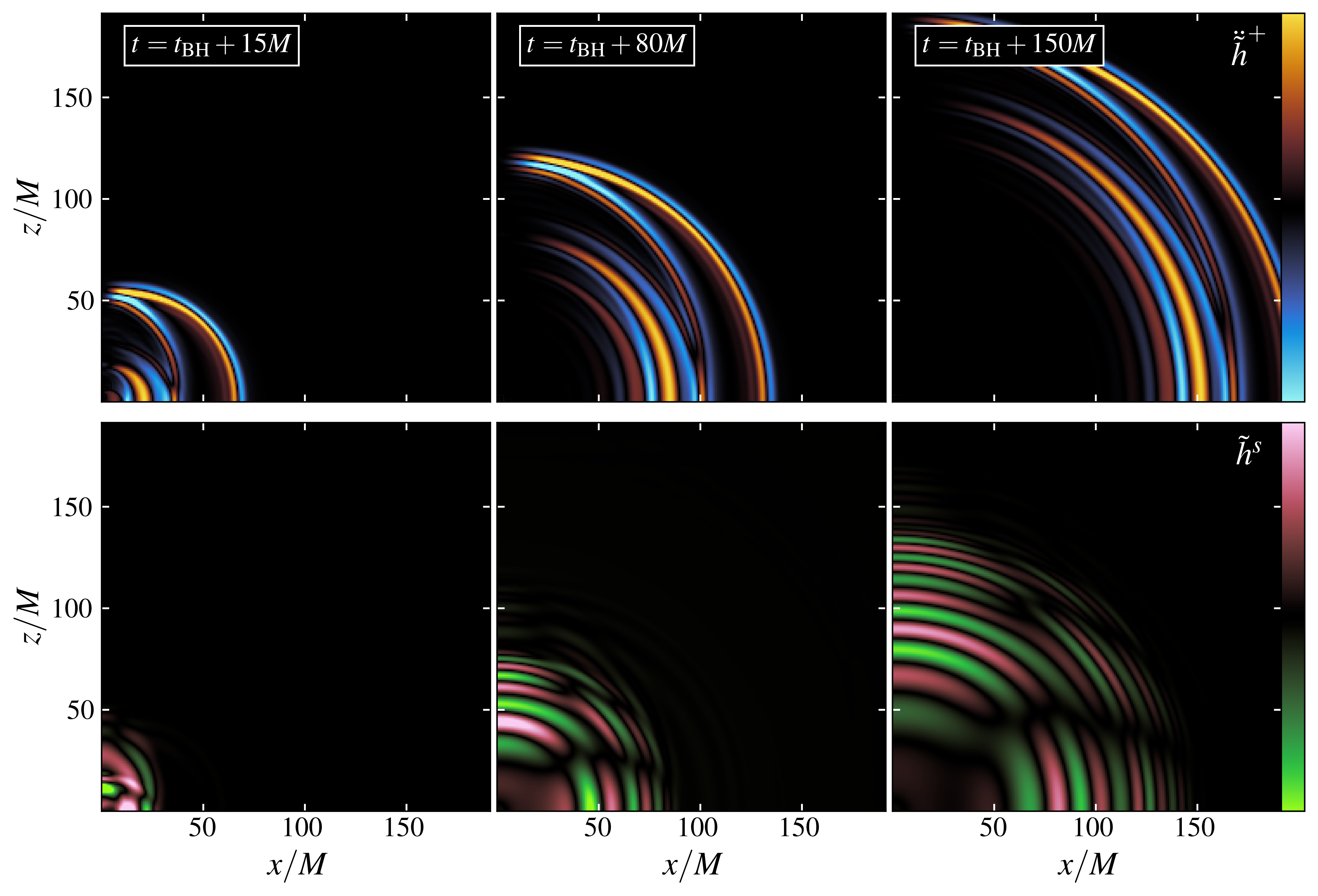}
        \caption{Tensor and scalar radiation spectrum comparison from the collapse of a circular loop with $(\eta\mpl^{-1},~R_0\mpl,~\alpha)=(0.04,~300,~200)$ to form a black hole at $t_\mathrm{BH}$. The figure shows the radiated gravitational waves from an edge-on point of view. The axial symmetry of the collapse only excites $(l,~m)=(\mathrm{even},~0)$ ``plus'' components being $(l,~m)=(2,~0)$ the dominant. The massive properties of the new scalar mode result in a much richer higher-mode spectrum with dispersive and sub-luminal properties. The first outer waves in the top panel correspond to junk radiation as a result of initial conditions, which occurs in numerical relativity simulations. A movie showing the collapse and radiation can be found \href{https://youtu.be/YYBwyAbH5Fk}{here} \cite{Movie2}.}
        \label{fig:radiation_spectrum}
    \end{figure*}

    In forecasting the ability of current and future gravitational-wave instruments to detect the presence of the scalar QNMs, one must know the initial amplitudes of the propagating scalar waves that are excited; if the amplitudes are too low, they are, obviously, undetectable. These amplitudes depend on the configuration that leads to the formation of the black hole. In the case of \textit{massive} scalar gravitational waves, not only are the triggered amplitudes important, but so is the propagation, which is far more complex than that of their massless counterparts. Even if they are excited, frequencies below the mass cutoff are damped away, and frequencies that do survive are dispersed and manifest as an \textit{inverse chirp} \cite{Sperhake:2017itk,Rosca-Mead:2019seq,Rosca-Mead:2020ehn}. A precise theoretical characterization of both their generation and propagation is essential for quantifying their impact.
    
    What might be the origin of these additional scalar profiles? The standard sources of tensor gravitational waves are binary black holes, but in standard scalar-tensor theories any initial hair is likely to have decayed away so one requires a dynamical mechanism to excite the scalar field during the merger \cite{Will:1989sk,Yunes:2011aa,Healy:2011ef,Silva:2017uqg,Figueras:2021abd}
    (for other interesting caveats, see \cite{Jacobson:1999vr,Alexander:2009tp,Horbatsch:2011ye,Sotiriou:2013qea}). Rotating neutron stars \cite{Doneva:2016xmf,Yazadjiev:2016pcb,Doneva:2018ouu} and neutron star binaries can support non-trivial scalar field profiles \cite{Damour:1993hw,Palenzuela:2013hsa,Ramazanoglu:2016kul,Andreou:2019ikc}, so they provide a possible source, along with other more exotic possibilities such as boson stars or topological defects. The latter objects may be endowed with internal self forces which, during gravitational collapse, enable and amplify the radiation of scalar tensor gravitational waves.
    
    There has been a detailed analysis of the scalar wave signals that emerge from spherically symmetric collapses \cite{Scheel:1994yr,Harada:1996wt,Novak:1997hw,Novak:1998rk,Novak:1999jg,Gerosa:2016fri,Rosca-Mead:2020ehn, Geng:2020slq}, showing the main properties of the scalar waves, which differ greatly from the familiar, and remarkably simple, propagation of tensor waves.
    In this paper we extend the previous work to study the generation of such scalar waves in systems that go beyond spherical symmetry. This allows us to compare the tensor gravitational waves (absent in spherically symmetric configurations) to their scalar counterparts.  We choose to study a toy model where the scalar source is exotic, arising from a topological defect -- the collapse of a non-minimally coupled cosmic string loop. In this case, the inherent tension of the string forces the configuration to collapse towards a single point -- the center of the loop. We show that this is a remarkably rich system with multi-mode tensor and scalar gravitational waves -- as well as decoupled (Goldstone) massless scalar waves, which we do not study in this paper. 
    
    Using numerical relativity simulations of this phenomenon, we aim to understand in more detail how the various waves are generated and propagate. We confirm the standard $1/r$ decay and luminal propagation of the tensor waves. The additional scalar mode, on the other hand, has a much richer behaviour. We construct the frequency spectrograms, which feature the novel \textit{inverse chirp} behaviour as described in \cite{Sperhake:2017itk,Rosca-Mead:2019seq,Rosca-Mead:2020ehn, Geng:2020slq}. However, the scalar waveforms extracted at finite distance do not show clearly the QNMs; the dispersive nature of the scalar mode obscures the ringdown phase, mixing different stages of the collapse. In order to alleviate the dispersion and scrambling of the extracted signals, we \textit{rewind} them using an effective massive wave equation in flat space. This technique allows us to classify the different stages of the collapse and clearly identify the scalar ringdown phase -- extracting the amplitudes and complex frequencies, which show excellent agreement with those expected from perturbative calculations \cite{Kokkotas:1999bd,Konoplya:2004wg,Konoplya:2006br,Dolan:2009nk,Tattersall:2018nve}. The fact that such a naive method of reconstructing the QNMs will be successful beyond spherical symmetry is not at all obvious, and therefore this is a useful result for those interesting in reconstructing initial amplitudes from extracted data.
    
    The outline of this paper is as follows: In Section \ref{sec:multiscalar} we present the formalism for multi-scalar tensor theories and the transformations between the Einstein and Jordan frames. In Section  \ref{sec:topdef} we construct the exotica -- string-like topological defects in the gravitational sector and describe their main properties. In Section \ref{sec:propwaves} we study the rich massless and massive radiation spectrum of the event (Fig. \ref{fig:radiation_spectrum}), and discuss the propagation properties of the new scalar waves. In Section \ref{sec:qnm}, we explain the rewinding technique used to reconstruct the ringdown waveforms. This also allows us to extract both the complex frequencies and amplitudes of the QNMs. Lastly, we conclude and suggest further research directions in Section \ref{sec:discussion}.

    \section{Multi-scalar-tensor theories}\label{sec:multiscalar}
    
    Consider a theory of a complex scalar field $\Phi$ non-minimally coupled to gravity with the following action in the Jordan frame
    \begin{align}
    S = \int d^4x \sqrt{-\tilde{g}}\biggl[ &F(\Phi,\Phi^*)\tilde{R} \nonumber\\
    &-\frac{1}{2}\tilde{g}^{\mu\nu}\nabla_\mu \Phi \nabla_\nu \Phi^*
    -V(\Phi,\Phi^*)\biggr]
    \label{eq:actionreal}
    \end{align}
    where $\Phi^*$ is the complex conjugate of $\Phi$ and $V(\Phi,\Phi^*)$ is a potential with a vacuum manifold which has a $U(1)$ symmetry
    \begin{align}
    V(\Phi, \Phi^*)=\frac{\lambda}{4}\left(\vert\Phi\vert^2-\eta^2\right)^2,
    \end{align}
    where $\lambda$ is the coupling constant and $\eta$ is the symmetry breaking scale. Although we can always expand the complex field into two real fields and then conformally transform \cite{Kaiser:2010ps}, sometimes it can be more convenient to work with a scalar field and its conjugate. We can treat them as a set of independent fields $\Psi^a=(\Phi, \Phi^*)$  so that the action is compactly written in the Einstein frame as
    \begin{align}\label{eq:act_einst}
    S = \int d^4 x \sqrt{-g}\biggl[\frac{\mpl^2}{2}R-\frac{1}{2}G_{ab}g^{\mu\nu}\nabla_\mu \Psi^a \nabla_\nu \Psi^b -\hat{V}\biggr],
    \end{align}
    where we now have
    \begin{align}\label{eq:g_transform}
    \tilde{g}_{\mu\nu} = \frac{\mpl^2}{2F}g_{\mu\nu}~\qquad
    \hat{V} = \left(\frac{\mpl^2}{2F}\right)^2V ,
    \end{align}
    and $\mpl$ is the reduced Planck mass. The complex field-space metric is computed via
    \begin{align}
    G_{ab} = & \frac{\mpl^2}{2F}\left(\frac{\mathbb{J}_{2\times 2}-\delta_{ab}}{2}\right)+\frac{3}{2}\frac{\mpl^2}{F^2}\frac{\partial F}{\partial \Psi^a}\frac{\partial F}{\partial \Psi^b},
    \label{eq:Gijcomplex}
    \end{align}
    where $\mathbb{J}_{2\times 2}$ is a $2\times 2$ matrix of ones. We now make the choice of
    \begin{eqnarray}\label{eq:F_coupling}
    F(\Phi,\Phi^*)=\frac{1}{2}\left(\mpl^2-\frac{\alpha}{6}\vert\Phi\vert^2\right),
    \end{eqnarray}
    which provides a sufficiently non-trivial toy model for our study -- the exact form is not important for our results. The non-minimal coupling arises as the first non-trivial correction at the level of an effective field theory of scalar-tensor theories. Indeed, it has been shown \cite{Herranen:2015ima} that it is invariably generated in the effective action in the case where $V$ goes beyond the usual Klein Gordon potential, $m^2|\Phi|^2$. Note also that $\alpha=1$ corresponds to the special case of a conformal coupling; in that case, and in the absence of $\mpl^2$ and a mass term in the potential, the theory would be conformally invariant. With that choice of non-minimal coupling we have that the field-space metric is
    \begin{align}
    G_{ab}=\frac{\mpl^2}{8F^2}
    \begin{pmatrix}
     \frac{\alpha^2}{12}{\Phi}^{*2} & \mpl^2-\frac{\alpha}{6}\left(1-\frac{\alpha}{2}\right)\vert\Phi\vert^2\\ 
    \mpl^2-\frac{\alpha}{6}\left(1-\frac{\alpha}{2}\right)\vert\Phi\vert^2 &  \frac{\alpha^2}{12}\Phi^2
     \end{pmatrix}
    \label{eq:complexGij}
    \end{align}
    
    The gravitational field obeys the standard Einstein equations sourced by the new energy momentum tensor
    \begin{align}
    T_{\mu\nu}&=G_{ab}\nabla_\mu \Psi^a \nabla_\nu \Psi^b 
    - g_{\mu\nu}\left(\frac{1}{2}G_{ab}\nabla_\beta \Psi^a \nabla^\beta \Psi^b+\hat{V}\right)
    \label{eq:complexemtensor}
    \end{align}
    and the equation of motion for the scalar field is given by
    \begin{align}
    \nabla_\mu\nabla^\mu\Psi^a+\tilde{\Gamma}^a_{bc}
    \nabla_\mu\Psi^b\nabla^\mu\Psi^c - G^{ab}\frac{\partial\hat{V}}{\partial\Psi^b}=0,
    \label{eq:eomfield}
    \end{align}
    where $G^{ab}$ is the inverse of the field-space metric $G^{ac}G_{cb}=\delta^a_b$ and $\tilde{\Gamma}^a_{bc}$ are the field space Christoffel symbols
    \begin{align}\label{eq:field_chris}
    \tilde{\Gamma}^a_{bc}=\frac{1}{2}G^{ad}\left(\frac{\partial G_{bd}}{\partial\Psi^c}+\frac{\partial G_{cd}}{\partial\Psi^b}-\frac{\partial G_{bc}}{\partial\Psi^d}\right).
    \end{align}
    
    Expanding the equation of motion for the complex scalar field, 
    \begin{align}\label{eq:eom}
    0&=\nabla_\mu\nabla^\mu\Phi  +\frac{\alpha{\Phi^*}}
    {6M_\mathrm{Pl}^2-\alpha\vert\Phi\vert^2}
    \nabla_\mu\Phi\nabla^\mu\Phi
    \nonumber\\
    &+\frac{6\mpl^2\alpha^2\Phi}
    {\left(6\mpl^2-\alpha\vert\Phi\vert^2\right)
    \left(6\mpl^2-\left(1-\alpha\right)
    \alpha\vert\Phi\vert^2\right)}
    \nabla_\mu\Phi\nabla^\mu{\Phi^*}
    \nonumber\\
    &-\frac{6\lambda \mpl^2
    \left(6\mpl^2-\alpha\eta^2\right)
    \left(\vert\Phi\vert^2-\eta^2\right)\Phi}
    {\left(6\mpl^2-\alpha\vert\Phi\vert^2\right)\left(6\mpl^2-\left(1-\alpha\right)
    \alpha\vert\Phi\vert^2\right)},
    \end{align}
    which is a wave equation with derivative coupling terms and sourced by the scalar potential in Eqn. \eqref{eq:g_transform}.

    \section{The progenitor: A circular string loop}\label{sec:topdef}
    
    Topological defects in scalar fields are an intriguing aspect of classical field theory that have consistently led to interesting insights in various aspects of fundamental physics. 
    Cosmic strings \cite{Kibble:1976sj,Vilenkin:1981kz} are a particular example that naturally arise after a phase  transition  in  the  early  universe,  when the symmetry of the vacuum is broken. There exists a plethora of  models encompassing gauge and global strings, which have been studied analytically as well as numerically over many decades \cite{Hindmarsh:1994re,Shellard:1987bv,Matzner:1988qqj,Laguna:1989hn,Matsunami:2019fss,Drew:2019mzc,Vincent:1997cx,Hindmarsh:2017qff,Nambu:1969se,Goto:1971ce}. Only recently has the full gravitational behaviour of field theory cosmic strings been studied using numerical relativity \cite{Aurrekoetxea:2020tuw,Helfer:2018qgv}\footnote{See \cite{Blanco-Pillado:2018ael, Chernoff:2018evo,Blanco-Pillado:2019nto} for ongoing work estimating the smoothing of Nambu-Goto strings including their linearized gravitational backreaction.}, showing that gauge circular loops could collapse to black holes, emitting $\sim 2\%$ of their initial mass in gravitational waves. In the case of global strings, the theory possesses a global symmetry, in which a complex scalar field with a massless Goldstone boson introduces a long-range force. This radiation channel allows the string to emit energy in the form of Goldstone bosons, and a key question was whether this could prevent strings from collapsing to black holes \cite{Fort:1993zb}.

    In this work, we focus on the collapse of global $U(1)$ strings where the scalar field non-minimally couples to the metric. To understand their properties, it is often useful to look at the vacuum manifold of the (Einstein frame) scalar potential
    \begin{equation}
    \hat{V}(\vert\Phi\vert^2)=\frac{\mpl^4}{\left(M_\mathrm{Pl}^2-\frac{\alpha}{6}\vert\Phi\vert^2\right)^2}\frac{\lambda}{4}\left(\vert\Phi\vert^2-\eta^2\right)^2 .
    \label{eq:potential}
    \end{equation}
    For the case in which $\alpha=0$, we recover the standard Higgs potential with a $U(1)$ symmetry where the vacuum is at $\vert\Phi\vert=\eta$. Vacuum string-like defects correspond to cylindrical configurations of the field $\Phi(\sqrt{x^2+y^2},~z)$ with boundary conditions
    \begin{align}
    \Phi(\sqrt{x^2+y^2}\rightarrow 0,~z)&=0 ,\nonumber \\ 
    \Phi(\sqrt{x^2+y^2}\rightarrow \infty,~z)&=\eta .
    \label{eq:bcs}
    \end{align}
    
    An important quantity that sets the properties of the propagating scalar waves is the mass (which also sets the width of the strings), and can be found by expanding the equation of motion Eqn. \eqref{eq:eom} around the vacuum $\Phi=(\eta+\delta\varphi)e^{ i\theta}$. Isolating the coefficients of the linear terms in $\delta\varphi$ and $\theta$,
    \begin{align}\label{eq:mass}
    m^2_\varphi &\approx \frac{12 \mpl^2}{6\mpl^2 - (1-\alpha)\alpha\eta^2}\lambda\eta^2,\\
    m^2_\theta &= 0,
    \end{align}
    corresponding to a massive radial mode -- the dilaton --  and a massless Goldstone boson as usual. As we see, the mass of the radial mode now also depends on the strength of the coupling to the Ricci scalar $\alpha$; in the case of minimal coupling ($\alpha=0$) or a conformally invariant coupling ($\alpha=1$), we revert to the mass obtained for standard global strings. 
    
    Using these objects as a fiducial model for axisymmetric collapse, we extract the tensor and scalar gravitational waves from the collapse of an initially stationary circular string loop of radius $R_0=300\mpl^{-1}$. The parameters of the theory are the symmetry breaking scale $\eta=0.04\mpl$, coupling constant $\lambda=1$, and non-minimal coupling $\alpha=200$, which result in a scalar mass of $m_\varphi\approx 0.017$. The system is massive enough to form a black hole of mass $M=23.5\pm 0.2\mpl$, so that $m_\varphi M \approx 0.4$. In Fig. \ref{fig:radiation_spectrum} we plot the radiation spectrum that we study in detail below.

    \section{Propagation of scalar-tensor gravitational waves}\label{sec:propwaves}

    To see how the new scalar gravitational mode arises in the signal it is useful to recall that the Jordan and Einstein frame metrics are related via Eqn. \eqref{eq:g_transform},
    \begin{align}
    \tilde{g}_{\mu\nu} = \frac{\mpl^2}{\mpl^2-\frac{\alpha}{6}|\Phi|^2}g_{\mu\nu} = \frac{\mpl^2}{\mpl^2-\frac{\alpha}{6}\varphi^2}g_{\mu\nu},
    \end{align}
    where we have decomposed the two degrees of freedom of the complex scalar field as
    \begin{equation}
    \Phi=\varphi e^{i\theta},
    \end{equation}
    with both $\varphi$ and $\theta$ real -- we will dub $\varphi$ the {\it dilatonic} and $\theta$ the {\it Goldstone} part of the field. Expanding to first order $g_{\mu\nu} = g_{\mu\nu}^0+h_{\mu\nu}$ and $\varphi = \varphi_0 + \delta\varphi$ leads to the following relationship between the Jordan and Einstein frame perturbations of the fields
    \begin{align}\label{eq:gw_transform}
    \tilde{h}_{\mu\nu} &= \tilde{h}^{TT}_{\mu\nu} + g_{\mu\nu}^0 \tilde{h}^s \nonumber \\
    &=\frac{\mpl^2}{\mpl^2-\frac{\alpha}{6}\varphi_0^2}\left[h^{TT}_{\mu\nu}+g_{\mu\nu}^0\frac{\alpha \varphi_0}{3\left(\mpl^2-\frac{\alpha}{6}\varphi_0^2\right)}\delta\varphi \right],
    \end{align} 
    where we identify the transformations for the transverse-traceless tensor $\tilde{h}^{TT}_{\mu\nu}$ and the additional breathing mode $\tilde{h}^s$. Defining the ``plus'' $\tilde{h}^+ \equiv \tilde{h}^{TT}_{xx} = -\tilde{h}^{TT}_{yy}$  and ``cross'' polarisations  $\tilde{h}^\times \equiv \tilde{h}^{TT}_{xy} = \tilde{h}^{TT}_{yx}$, we obtain that we can relate the Jordan frame gravitational waves to quantities in the Einstein frame as
    \begin{align}\label{eq:tensorstrain}
        (\tilde{h}^+,~\tilde{h}^\times) &\equiv \frac{\mpl^2}{\mpl^2-\frac{\alpha}{6}\varphi_0^2} (h^+,~h^\times) \\
        \tilde{h}^s &\equiv \frac{\alpha \mpl^2 \varphi_0}{3\left(\mpl^2-\frac{\alpha}{6}\varphi_0^2\right)^2}\delta\varphi , \label{eq:scalarstrain}
    \end{align}
    which correspond to the three gravitational-wave degrees of freedom we have in this theory. In what follows, we further decompose each mode into spin -2 and spin 0 spherical harmonics
    \begin{align}
    \tilde{h}_{lm}^{+\times}(t) &= \int d\Omega~ \tilde{h}^{+\times}(t,\theta,\phi) ~\left[{}_{-2}Y_{lm}(\theta,\phi))\right]^*~, \\
    \tilde{h}_{lm}^s(t) &= \int d\Omega~ \tilde{h}^s(t,\theta,\phi) ~\left[{}_{0}Y_{lm}(\theta,\phi)\right]^* .\label{eq:decomp_scalar}
    \end{align}

    \subsection{Tensor gravitational waves}
    
    \begin{figure}[t]
        \includegraphics[width=\columnwidth]{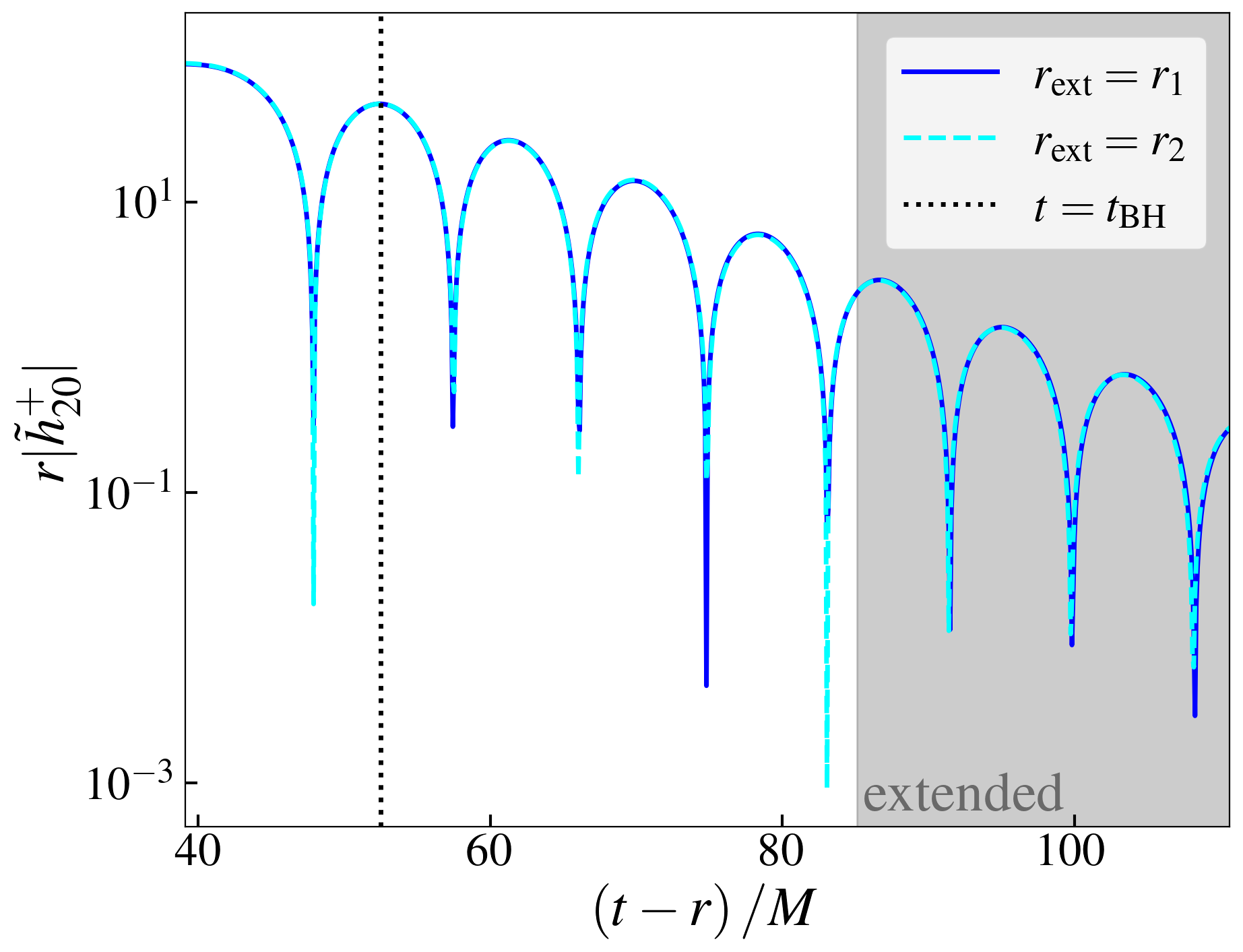}
        \caption{Dominant tensor $\tilde{h}^+_{20}$ waveform focusing on the ringdown phase of a black hole formed from a loop with $(\eta \mpl^{-1},~R_0\mpl,~\alpha)=(0.04,~300,~200)$. The vertical dashed line corresponds to the black hole formation time, defined as when $\ddot{h}^+_{20}$ peaks. Note that we have removed the numerical noise present at later times when the signal is subdominant by extending it using the extracted spin-2 $(l,~m)=(2,~0)$ QNM frequency $M\omega^+_{20} = 0.372 - \iu 0.089~ [\pm 0.004 - \iu 0.001]$, consistent with perturbative predictions: $M\omega^+_{20} = 0.3735 - \iu 0.0890$.
        }
        \label{fig:tensor_waveform}
    \end{figure}
    
    We have constructed the tensorial strain waveforms in the Jordan frame $(\tilde{h}^+,~\tilde{h}^\times)$ by integrating the Weyl scalar $\Psi_4$ \cite{Newman:1961qr}  extracted from our numerical simulations in the Einstein frame with the tetrads proposed by \cite{Baker:2001sf} 
    \begin{equation}
        \Psi_4 = \ddot{h} = -\ddot{h}^+ + \iu \ddot{h}^\times .
    \end{equation}
    We then transform this using Eqn. \eqref{eq:tensorstrain}. Given the symmetry of the collapse, we only excite $(l,~m)=(\mathrm{even},~0)$ modes of the ``plus" tensor $\tilde{h}^+_{lm}$ polarisation, with $(l,~m)=(2,~0)$ the dominant one. The signal features a low frequency infall, collapse during which the black hole forms and a ringdown. We show the propagation of the radiation in the top three panels of Fig. \ref{fig:radiation_spectrum}. In Fig. \ref{fig:tensor_waveform} we plot the ringdown phase of the signal and confirm the luminal propagation and expected $1/r$ decay by extracting the signal at two different radii $\rext=\lbrace 100M,~ 125M\rbrace $. The extracted waveform matches the expected QNM frequency prediction: $M\omega^+_{20} = 0.372 - \iu 0.089~ [\pm 0.004 - \iu 0.001]$ vs $M\omega^+_{20} = 0.3735 - \iu 0.0890$.

    \subsection{Scalar gravitational waves}

\begin{figure}[t]
        \includegraphics[width=\columnwidth]{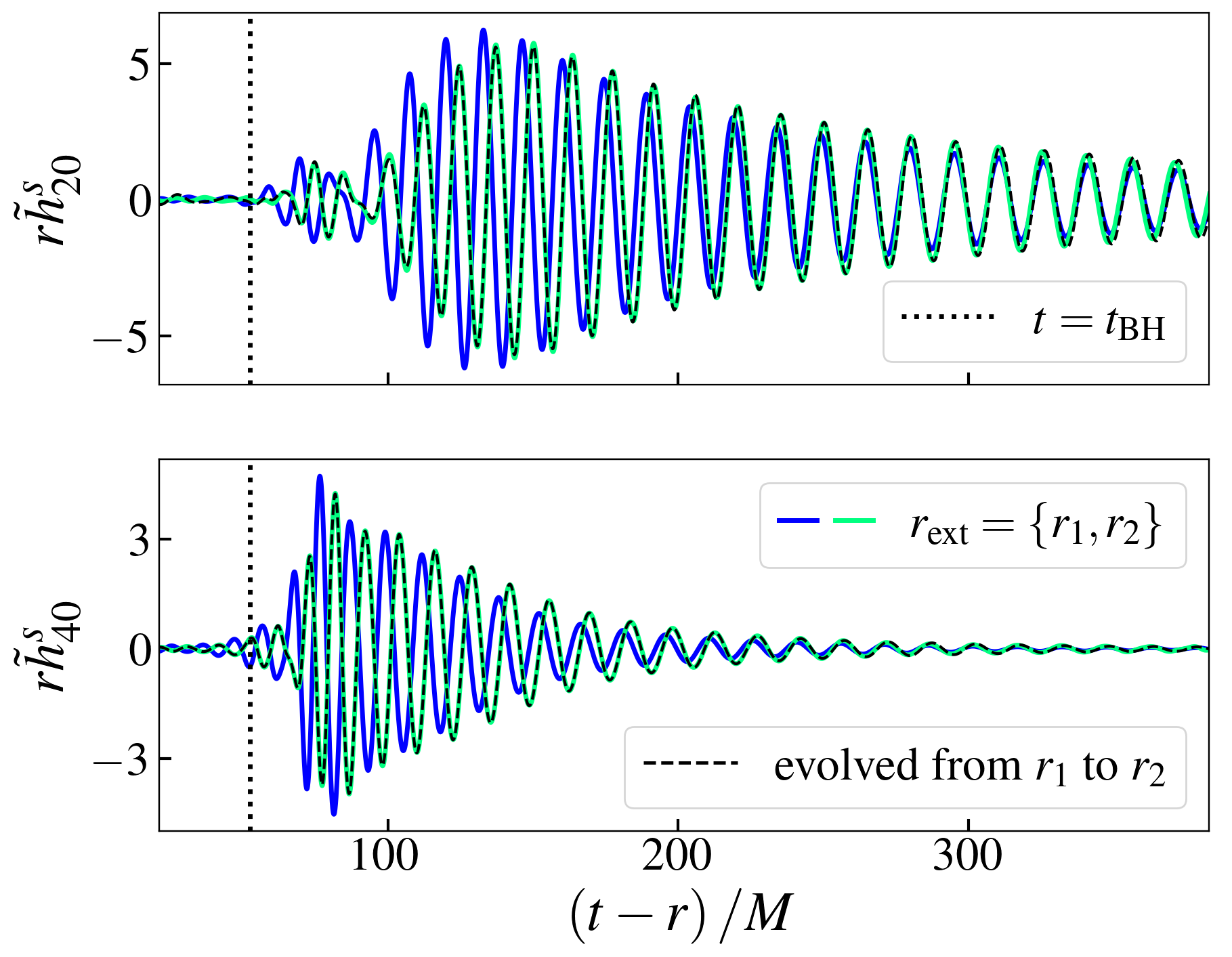}
        \caption{Scalar $\tilde{h}^s_{lm}$ waveforms constructed from the evolution of the massive dilaton $\varphi$. Top and bottom panels show the dominant extracted spin-0 harmonic waves, $(l,~m)=(2,~0)$ and $(l,~m)=(4,~0)$ respectively. Blue and green illustrate different extraction radii $\rext=\lbrace 100M,~ 125M\rbrace $ respectively, and the black dashed lines show the agreement with the larger radius extraction (green line) assuming a flat space massive wave equation.
        }
        \label{fig:dilaton_waveform}
    \end{figure}

    Asymptotically  far  from  the  black  hole  (or  the  collapse event), the scalar field rests at the vacuum, set by the symmetry breaking scale $\varphi_0=\eta$. The scalar waves that are produced during and post-collapse travel to infinity and cause the field to oscillate around the symmetry breaking scale $\varphi=\eta+\delta\varphi$. We construct the scalar strain waveforms $\tilde{h}^s$ from the evolution of these perturbations $\delta\varphi = \varphi - \eta$ using Eqn. \eqref{eq:scalarstrain}. The propagation of the scalar waves is shown in the bottom three panels of Fig. \ref{fig:radiation_spectrum}. We now decompose the extracted signals into spin-0 weighted spherical harmonics $\tilde{h}^s_{lm}$, Eqn. \eqref{eq:decomp_scalar}, and find that given the symmetry of the collapse, only $(l,~m)=(\mathrm{even},~0)$ modes are excited. In Fig. \ref{fig:dilaton_waveform} we plot the loudest modes\footnote{We choose the parameters such that the lowest $(l,~m)=(0,~0)$ is a \textit{non-propagating} wave.} for extraction radii $\rext=\lbrace 100M,~ 125M\rbrace $. When overplotting the waveforms, we find that massive scalar waves experience a faster than $1/r$ decay as expected, together with a delay due to their subluminal propagation speeds.

    The additional scalar degree of freedom evolves according to the Klein-Gordon equation in a Schwarzschild background,
    \begin{equation}\label{eq:curved_KG}
        \left[\partial_t^2-\partial_{r_*}^2+V_s(r)\right] r\tilde{h}_{l}^s=0 ,
    \end{equation}
    with $r_* \equiv r + 2M\ln\left(r-2M\right)$ the tortoise coordinate and $V_s(r)$ the effective potential
    \begin{equation}
        V_s(r) = \left(1-\frac{2M}{r}\right)\left(m_\varphi^2+\frac{l(l+1)}{r^2}+\frac{2M}{r^3}\right) ,
    \end{equation}
    where $m_\varphi^2$ is given in terms of the parameters of the theory, Eqn. \eqref{eq:mass}. Far from the source, $V_s(r\rightarrow \infty)\approx m_\varphi^2$ and $r_*\rightarrow r$, thus the propagation reduces to a massive Klein-Gordon equation in flat space. So, given a plane wave of the form $\exp{[-\iu(\omega t+k_i x^i)]}$, where $\omega$ and $k^i$ are the frequency and 3-momentum, the scalar waves will have a dispersion relation $\omega^2=k^2+m_\varphi^2$.\\
    
    \textit{Non-propagating waves}  have $\omega^2\le m_\varphi^2$ and a spatial dependence of the form $e^{-|k|r}$. Note that the wave-numbers are complex\footnote{These are sometimes called \emph{evanescent waves}. See \cite{Golat:2019aap} for an application to gravitational waves.} and we recover, in the static case ($\omega=0$), the usual Yukawa potential with $h^s\propto e^{-m_\varphi r}$. This is confirmed in our simulations by looking at the Fourier transform of the extracted wave, where frequencies below $\omega^2<m_\varphi^2$ are exponentially suppressed. 
    
\begin{figure}[t]
        \includegraphics[width=\columnwidth]{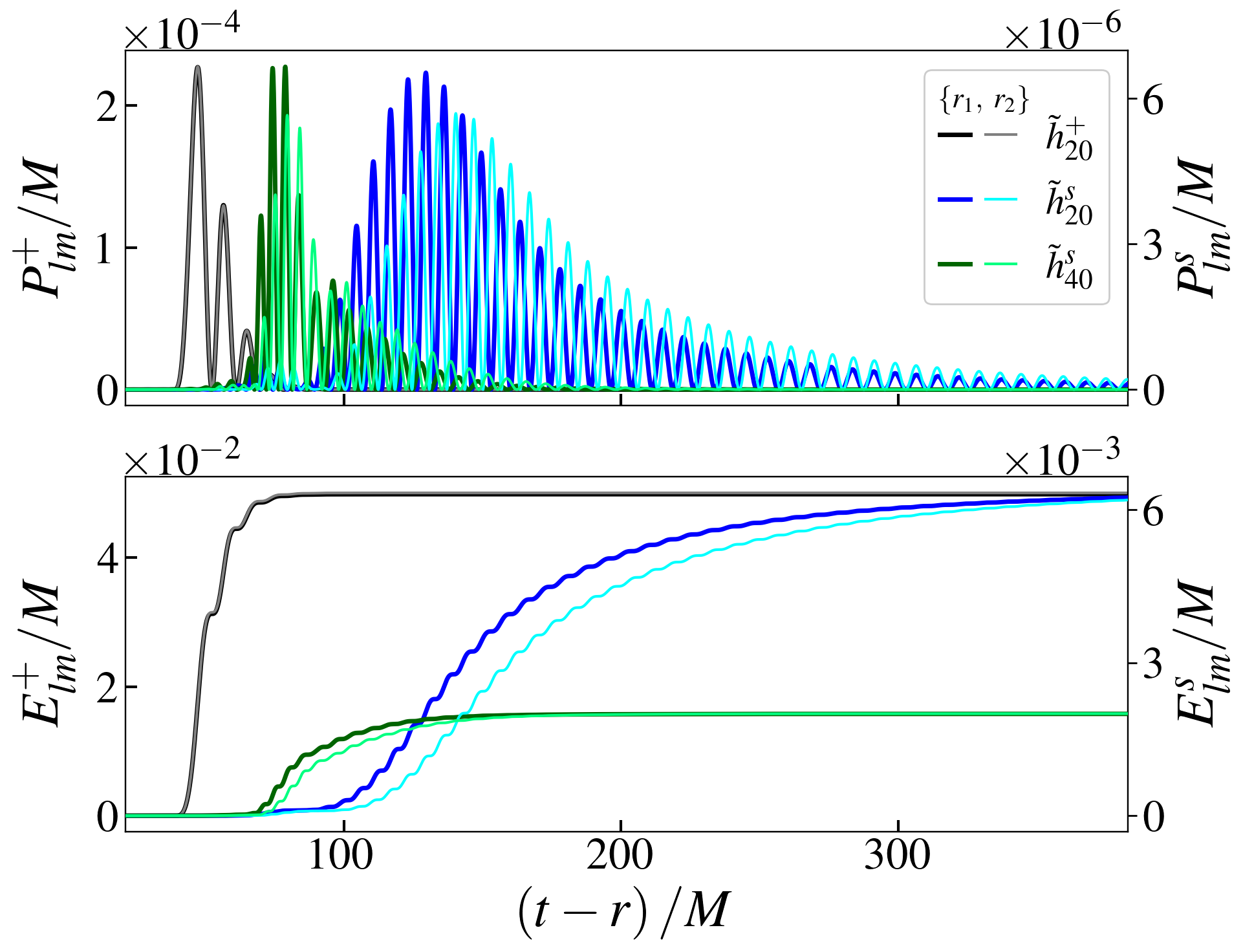}
        \vspace{-1.5mm}
        \caption{Power and energy contained in the tensor and scalar waveforms via Eqn. \eqref{eq:power_scalar} for extraction radii $\rext=\lbrace 100M,~ 125M\rbrace $. The luminosity of the scalar waves decays faster than $1/r^2$, but the integrated energy remains constant as the signal spreads out. Even though tensor modes are $>40$ times louder for this extraction radius, the amount of energy emitted in scalar waves is only $6$ times smaller.}
        \label{fig:dilaton_powerenergy}
    \end{figure}

    \textit{Propagating waves}, on the other hand, satisfy $\omega^2> m_\varphi^2$, and the general solution is a superposition of plane waves of the form $\exp{[-i(\omega t+k_i x^i)]}$. If we now consider a massive wave packet emitted by an event or source, we can see how it will differ from the standard massless tensor waves. For a start, its group velocity, 
    \begin{equation}\label{eq:group_v}
    v_g\equiv \frac{\partial \omega}{\partial k}=\sqrt{\frac{k^2}{k^2+m_\varphi^2}}=
    \sqrt{1-\left(\frac{m_\varphi}{\omega}\right)^2}
    \end{equation}
    shows the the process will be dispersive. Lower frequency waves will propagate more slowly and dissipate. We check these properties in the top panel of Fig. \ref{fig:dilaton_powerenergy}, where we observe a delay in the arrival of the power peak between different frequency modes. This $k$ dependence of the propagation results in the initial waveform being substantially deformed as it propagates from $r\approx 2M$ to $r_\mathrm{ext}$, scrambling the shape of the wave. Higher frequency waves propagate at higher group velocity reaching the detector first and resulting in a generic \textit{inverse chirp} signal, as shown in \cite{Sperhake:2017itk,Rosca-Mead:2019seq,Rosca-Mead:2020ehn}. This means that, unlike in the case of the  tensor sector, information about the source event (in our case, the cosmic string collapse) may be lost in the process. These two effects -- the faster decay in peak amplitude and the deformation of the signal -- make it potentially difficult to reconstruct the massive scalar mode from a gravitational waveform at larger distances.

    We test our results by evolving the $\rext=100M$ extracted signal to $r=125M$ using the Fourier techniques introduced in \cite{Rosca-Mead:2020ehn}. The authors describe the outgoing scalar wave related to the numerically extracted signal as
    \begin{align}
        r\tilde{h}(t;r)=\int\frac{d\omega}{2\pi} &\mathcal{F}\lbrace r\tilde{h}(t;r)\rbrace e^{-i\omega t} ,
    \end{align}
    where $k^+ = +\sqrt{\omega^2-m_\varphi^2}$. The function $\mathcal{F}\lbrace r\tilde{h}(t;r)\rbrace$ is the Fourier transform at the target radius and is related to the one extracted from simulations via
    \begin{align}
        \mathcal{F}\lbrace r\tilde{h}(t;r)\rbrace=\mathcal{F}\lbrace r\tilde{h}(t;&r_\mathrm{ext})\rbrace \times \nonumber\\
        &\begin{cases}
            e^{-ik^+\left(r-r_\mathrm{ext}\right)}, & \text{if } \omega\leq -m_\varphi\\
            e^{+ik^+\left(r-r_\mathrm{ext}\right)}, & \text{if } \omega> -m_\varphi
            \end{cases}\label{eq:outwards}
    \end{align}
    We plot the results in Fig. \ref{fig:dilaton_waveform}, showing excellent agreement with the evolution of a massive wave equation in flat space.
    
    As a result of the dispersive nature of massive waves, the luminosity of such a signal decays more quickly than the $1/r^2$ we obtain for massless fields. However, energy is conserved; the decay in amplitude of the waveform is compensated by the growth in the width of the wave packet. We confirm this in Fig. \ref{fig:dilaton_powerenergy}, computing the power and energy in the scalar and tensor sector for $r_\mathrm{ext}=\lbrace 100M,~ 125M\rbrace$,
    \begin{align}
    P^{+\vert s}_{lm}(t) & = \frac{\mathrm{d} E^{+\vert s}_{lm}}{\mathrm{d} t} \propto \frac{r^2}{16\pi \tilde{G}} \left(\frac{\partial \tilde{h}^{+\vert s}_{lm}}{\partial t}\right)^2 , \label{eq:power_scalar}
    \end{align}
    with $\tilde{G} \equiv 1/F(\varphi=\eta)$ the effective Newton's constant via Eqn. \eqref{eq:F_coupling}. In the top panel of Fig. \ref{fig:dilaton_powerenergy}, we see that at these ``astronomically close'' extraction radii, the scalar gravitational waves are already $40$ times fainter, and they will become more subdominant as they propagate. However, the bottom panel shows that the energy contained in the scalar waves -- which stays constant -- is only $6$ times smaller, which could still indirectly impact the tensor waveforms, e.g. by changing their phase evolution relative to the GR case \cite{Berti:2018cxi, Silva:2022srr,Nair:2019iur,Maselli:2020zgv,Barsanti:2022ana}.
    
    We often want to extrapolate the signals extracted from simulations to what an asymptotic observer would see. Tensor gravitational waves, which decay as $1/r$ and propagate at the speed of light are in general easily extended by extracting at several different radii \cite{Bishop:2016lgv,Iozzo:2020jcu}\footnote{See \cite{Bishop:1996gt,Winicour:2008vpn,Babiuc:2011qi,Moxon:2021gbv} for alternative waveform extraction methods.}. The dispersive properties of massive waves, on the other hand, make this process a harder challenge. However, it has been shown that in the large distance limit the \textit{stationary phase approximation} can be used to construct the asymptotic waveforms of the propagating scalar wave \cite{Rosca-Mead:2020ehn}.
    
    \begin{figure*}[t]
        \includegraphics[width=\linewidth]{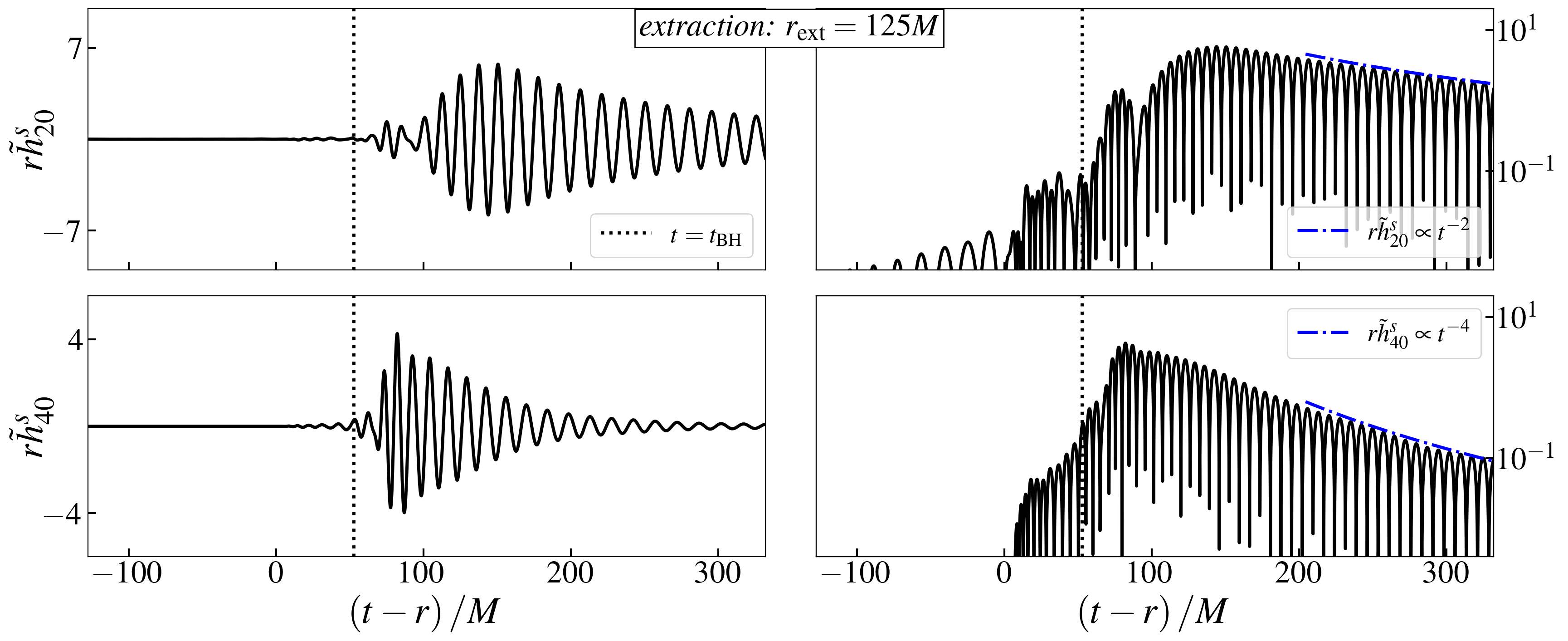}
        \caption{Scalar waveforms extracted at $\rext=125M$. Top and bottom panels plot the $(l,~m)=(2,~0)$ and $(l,~m)=(4,~0)$ modes (the right panel plots show the absolute value of the waveforms). The vertical dashed line illustrates the time at which the black hole forms -- defined as the peak of the tensor waveform. We find no evidence of exponentially decaying waves corresponding to the QNMs. In addition, there is a delay between the peak amplitudes of the $h^s_{20}$, $h^s_{40}$ and $h^+_{20}$ modes, caused by the frequency dependent group velocity of scalar waves, Eqn. \eqref{eq:group_v}. On the right panel we show the late time behaviour of the signals, with $t^{-2}$ and $t^{-4}$ tail decays, consistent within the early-time $t^{-l-3/2}$ and late-time $t^{-5/6}$ power-law behaviours \cite{Hod:1998ra,Koyama:2001ee,Koyama:2001qw}.}
        \label{fig:dilaton_ext}
    \end{figure*}
    
    \section{Quasinormal mode ringdown phase}\label{sec:qnm}

    When black holes form and ring down to the static Schwarzschild solution, they radiate energy in the form of waves with a characteristic set of complex frequencies determined by the quasinormal modes. To do a precise study of the ringdown frequencies one has to decide the relevant time-window for such an analysis. In Fig. \ref{fig:tensor_waveform} we focused our attention on the post-BH formation phase of the tensor modes, where the signal shows an exponential decay. Given the luminal propagation of massless fields, this translates to defining the correct start time of QNMs, which is a well-known open problem \cite{Finch:2021qph}.  In Fig. \ref{fig:tensor_waveform} we took the start of the QNMs to be $t_0-t_\mathrm{peak}\approx 15M$ and we confirmed that the extracted frequencies agree with those values obtained from perturbative calculations \cite{Kokkotas:1999bd,Nollert:1999ji,Berti:2009kk}
    \begin{equation}
        M\omega^+_{20} = 0.372 - \iu 0.089~ [\pm 0.004 - \iu 0.001]
    \end{equation}

    For the scalar gravitational waves, on the other hand, the challenge is greater. Given their dispersive properties, different frequency components emitted at different stages of the collapse invariably mix. In fact, in Fig. \ref{fig:dilaton_ext} we find no evidence of exponentially decaying modes. Furthermore, unlike massless GW, scalar waves sourced during the pre-ringdown phases mixes with that of post-collapse ringdown, rendering the determination of the start of ringdown $t_0$ difficult.  
    
\begin{figure*}[t!]
        \includegraphics[width=\linewidth]{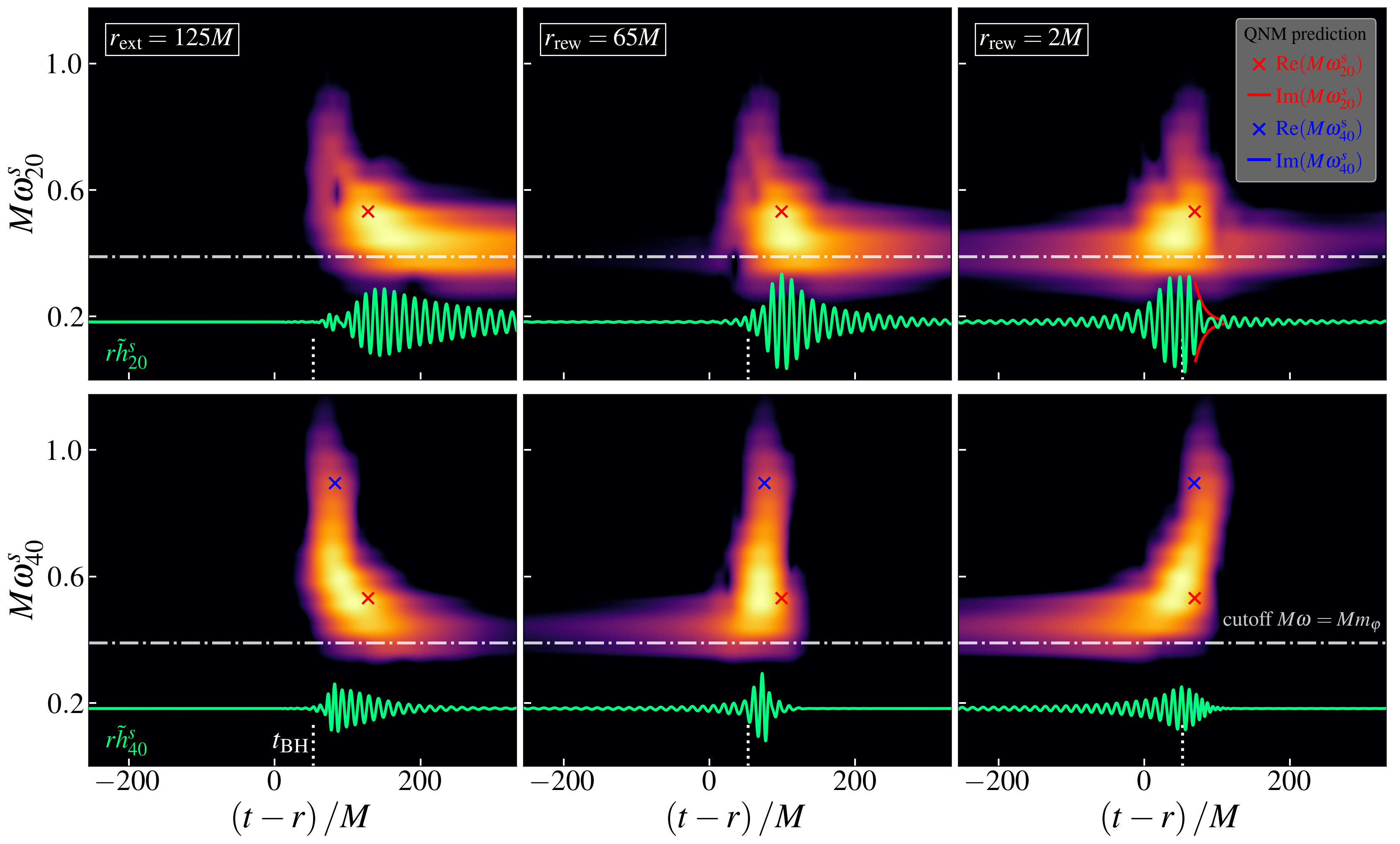}
        \caption{Spectrogram of the additional scalar modes extracted at $\rext=125M$ (left panels), rewounded to $\rew=65M$ (center panels) and $\rew=2M$ (right panels).  For reference we indicate the time of BH formation, $t_\mathrm{BH}$, defined as when $\ddot{h}^+_{20}$ peaks. In the left panel, the spectrogram shows an \textit{inverse chirp} behaviour due to the dispersive propagation of the massive scalar waves. The spectrogram crosses through the scalar QNM predictions $\mathrm{Re}(M\omega^s_{20})=0.533$ and $\mathrm{Re}(M\omega^s_{40})=0.896$, but without evidence of an exponentially decaying ringdown phase in the waveforms. In the central and right panels, we have evolved the extracted signal backwards to smaller radii using an effective massive wave equation, Eqn. \eqref{eq:eff_KG}. We recover a standard chirp waveform with real and imaginary QNM frequencies, which agree with perturbative calculations. A movie showing the rewinding can be found \href{https://youtu.be/Q3K9TP1effE}{here} \cite{Movie1}.}
        \label{fig:dilaton_spectrogram}
    \end{figure*}

    To attack this problem, we study the frequency content of the full waveform; if QNMs have been excited and are loud enough then they will be picked up in the analysis. We computed the spectrograms from the $\tilde{h}^s_{20}$ and $\tilde{h}^s_{40}$ waveforms extracted at $r=125M$. Even though we do not observe a clear exponentially decaying ringdown phase, the inverse chirp contains frequency contributions from the QNM predictions (red and blue crosses in the left panels of Fig. \ref{fig:dilaton_spectrogram}). This is because the ringdown phase sourced when the black hole forms has been obscured when mixing with louder stages of the collapse. We will then ``clean'' the extracted signals to have access to the different stages of the collapse as follows.

    The signals we extract at $\rext$ have been dispersed when propagating from $r\approx 2M$ to $\rext$; that is, from the source of the event to the extracted radius. During this process, the  amplitudes of the scalar gravitational waves are suppressed  below the mass cutoff scale $\vert\omega\vert < m_\varphi$, which is completely lost by the time we extract the signal at $\rext$. Meanwhile, waves with frequencies $|\omega| > m_{\varphi}$ survive, and  has been mixed in the time-domain waveforms during their propagation to the extraction zone. To ``unmix'' them, we propagate these scalar waves from $\rext$ back to the vicinity of the black hole at $\rew \approx 2M$ --  we \textit{rewind} the signals in order to remove the effect of the dispersion. During most of the backward propagation, the evolution is well described by the same massive flat space wave equation in spherical coordinates. At some point, the curvature and spherical harmonic terms will become relevant; but for simplicity, instead of solving Eqn. \eqref{eq:curved_KG},  we approximate the rewind of the scalar mode as an effective flat space wave equation\footnote{Ideally, one would like to rewind using Eqn. \eqref{eq:curved_KG}, which encodes the full description of the propagation, but this is a much harder computational task.}
    \begin{equation}\label{eq:eff_KG}
        \left[\partial_t^2-\partial_{r_*}^2+m^2_\mathrm{eff}\right] r\tilde{h}_{l}^s=0~,
    \end{equation}
    with a space independent effective mass $m_\mathrm{eff}^2$ that we define as the averaged value of the effective potential $V_s$ 
    \begin{align}
        m_\mathrm{eff}^2 =& \frac{\int_{\rext}^{\rew} V_s(r) dr}{\rew - \rext} = \nonumber \\
        =&m_\varphi^2 +\frac{2Mm_\varphi^2}{\rew-\rext}\log \frac{\rext}{\rew} + \frac{l(l+1)}{\rew\rext}\left(1-M\frac{\rew+\rext}{\rew\rext}\right) \nonumber\\
        &+M\left(\frac{\rew+\rext}{\rew^2\rext^2}\right) - \frac{4M^2}{3}
        \left(\frac{\rew^2+\rew\rext+\rext^2}{\rew^3\rext^3}\right)
    \end{align}
    So, for example, if we want to rewind a signal extracted at $\rext=125M$ to $\rew=2M$, 
    the effective mass describing such a propagation will be given by
    \begin{align}
        \frac{m_\mathrm{eff}^2}{m_\varphi^2} \approx 0.9327  +\frac{0.02602^2}{m_\varphi^2M^2}\left(1 + 2.907~l(l+1) \right)~.
    \end{align}
    In these simulations $m_\varphi M \approx 0.4$, so for the $l=2$ and $l=4$ modes studied in this work, we get $m_\mathrm{eff}^2 \approx 1.01 m_\varphi^2$ and $m_\mathrm{eff}^2 \approx 1.19m_\varphi^2$, respectively.\\

    \begin{figure*}[t]
        \includegraphics[width=\linewidth]{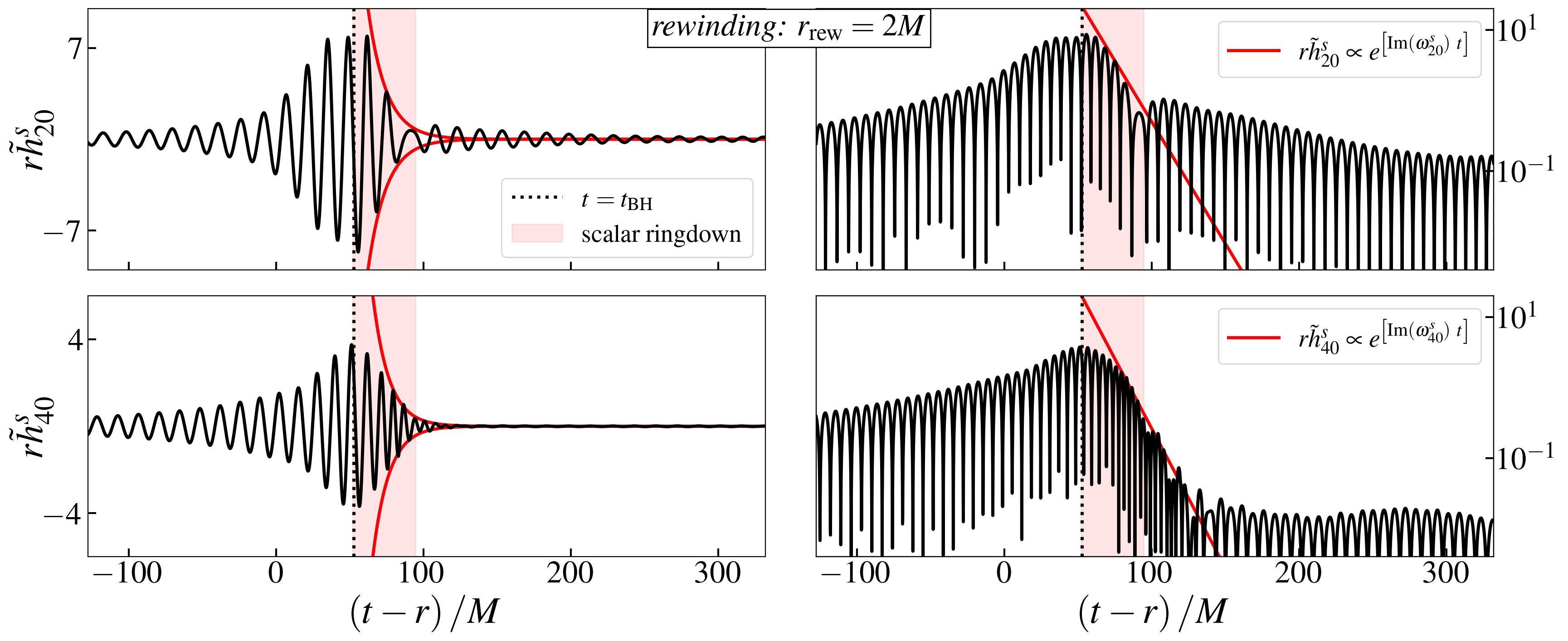}
        \caption{Scalar waveforms after the rewinding to $\rew=2M$ to alleviate the dispersion that has happened when propagating until extracted at $\rext=125M$ (the right panel plots show the absolute value of the waveforms). As opposed to Fig. \ref{fig:dilaton_ext}, we now observe a time-range compatible with the tensor ringdown where both scalar modes exponentially decay at the rate consistent with QNM calculations $\mathrm{Im}(M\omega^s_{20})= -0.0796$ and $\mathrm{Im}(M\omega^s_{40})= -0.0910$, see Ref. \cite{Tattersall:2018nve}. We identify this as the scalar ringdown, which is followed by a phase where remnants of the scalar field around the black hole source more scalar waves.}
        \label{fig:dilaton_inv}
    \end{figure*}
    
    Similar to outgoing propagation, we can solve this using Fourier techniques \cite{Rosca-Mead:2020ehn}
    \begin{align}
        r\tilde{h}(t;r)=\int\frac{d\omega}{2\pi} &\mathcal{F}\lbrace r\tilde{h}(t;r)\rbrace e^{-i\omega t}.
    \end{align}
    To study the rewinding proccess, we modify Eqn. \eqref{eq:outwards} to
    \begin{align}
        \mathcal{F}\lbrace r\tilde{h}(t;r)\rbrace=\mathcal{F}\lbrace r\tilde{h}(t;&r_\mathrm{ext})\rbrace \times \nonumber\\
        &\begin{cases}
            e^{-ik^+\left(r-r_\mathrm{ext}\right)}, & \text{if } \omega\leq m_\varphi\\
            e^{+ik^+\left(r-r_\mathrm{ext}\right)}, & \text{if } \omega> m_\varphi
            \end{cases}
    \end{align}
    We note that this technique allows us to rewind the extracted signals, but not to recover those non-propagating frequencies $\vert\omega\vert<m_\varphi$ that have been exponentially suppressed and never reach $\rext$.

    In Fig. \ref{fig:dilaton_spectrogram} we plot the extracted waveforms at $\rext = 125M$ (left panels), as well as the rewound signals to $\rew=\lbrace 65M,~ 2M\rbrace$ (center and right panels). This plot aims to illustrate how the inverse chirp is reversed as the signals are rewounded, recovering the standard chirp waveforms for $\rew = 2M$. In addition, we note that the peak of the scalar signals converges to black hole formation time. We therefore focus on the $\rew = 2M$ rewounded signal plotted in Fig. \ref{fig:dilaton_inv}, where we can now clearly distinguish three main stages of the collapse. First, the \textit{pre-BH formation} phase features an oscillatory behaviour of frequency $\omega\approx m_\varphi$ with increasing amplitude. The system radiates scalar gravitational waves of the scalar mass frequency as it collapses. The amplitude of the signal peaks at the BH formation time, consistent with the tensor sector. This is followed by the \textit{ringdown} phase (red-shaded region), characterised by a set of exponentially decaying QNMs, before they become subdominant at the \textit{post-BH formation} phase, when the remaining scalar field around the black hole mainly sources the scalar waves. This is analogous to the expected post-merger tensor waveforms in systems where there is remnant matter around the formed black hole -- such as neutron or boson stars mergers.\\

\begin{figure}[t]
        \includegraphics[width=\columnwidth]{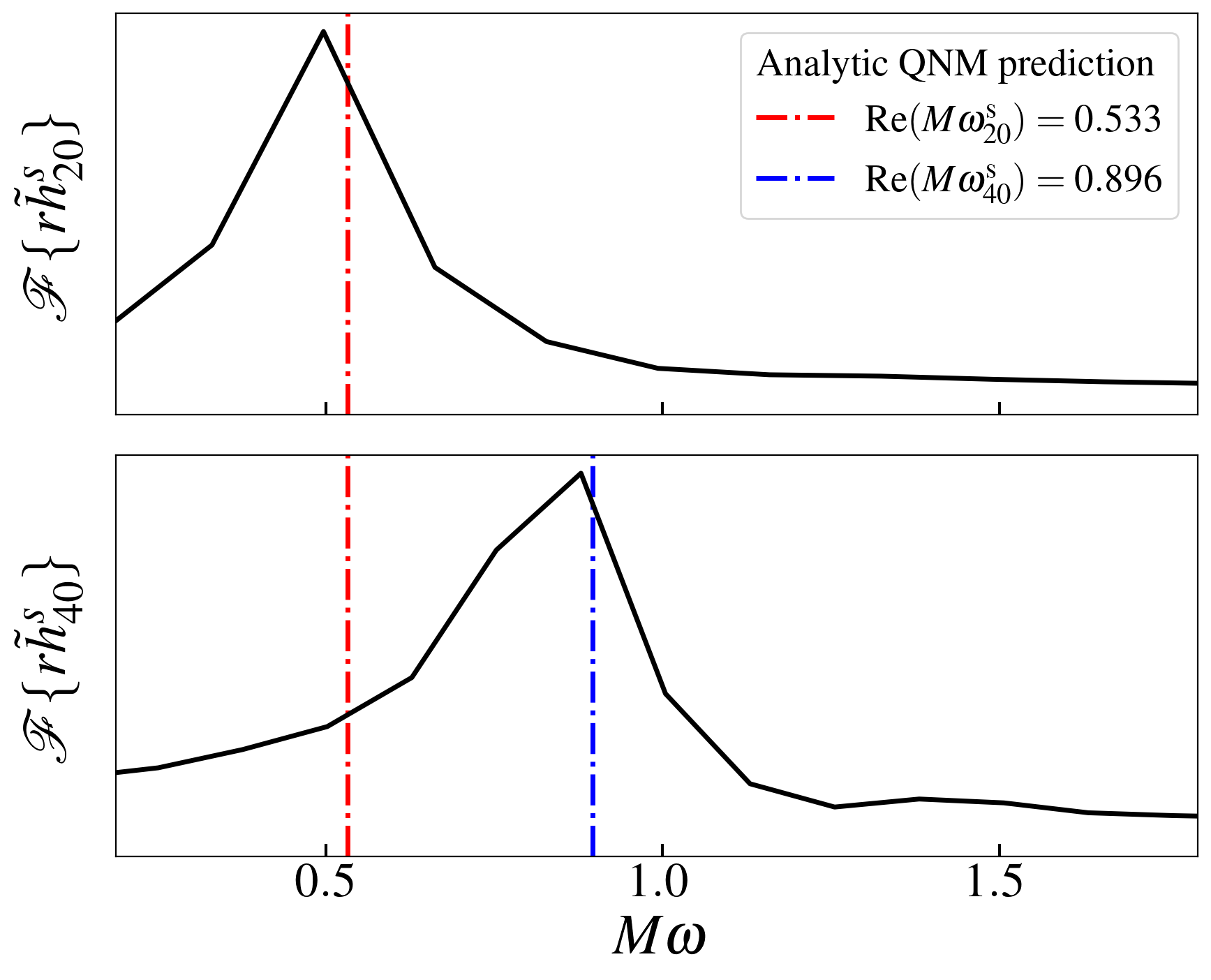}
        \caption{Real part of the scalar QNM frequencies extracted by Fourier transforming the range corresponding to the scalar ringdown phase in Fig. \ref{fig:dilaton_inv}, showing agreement with QNM predictions from Ref. \cite{Tattersall:2018nve}.
        }
        \label{fig:QNM_frequencies}
    \end{figure}

     We define the ringdown phase as the approximate time range where a clear exponential decay is observed. Therefore, we now focus the analysis on the region where the exponential decay provides the best fit for each of the modes, that is: $t/M\approx 60 - 100$ for $h^+_{20}$ and $t/M\approx 70 - 120$ for $h^+_{40}$. We first compute the analytical QNM frequency predictions following the expansion methods described in \cite{Tattersall:2018nve},
    \begin{align}\label{eq:qnm_freqs}
        M\omega^s_{20} &= 0.5326-\iu 0.0796~, \\
        M\omega^s_{40} &= 0.8960-\iu 0.0910~.
    \end{align}
    In Fig. \ref{fig:dilaton_inv} we plot the expected decay rates given by the imaginary part of the QNM frequencies, and show excellent agreement within the data in the scalar ringdown phase. In order to test the real part of the QNM frequencies, we construct the corresponding Fourier transforms in Fig. \ref{fig:QNM_frequencies} and compare where they peak
    \begin{align}
        \mathrm{Re}(M\omega^s_{20}) = 0.50 \pm 0.09~, \\
        \mathrm{Re}(M\omega^s_{40}) = 0.88 \pm 0.10~,
    \end{align}
    to the analytical predictions (dash-dotted vertical lines), also showing excellent agreement.

    \section{Discussion}\label{sec:discussion}

    In this paper we have studied the generation and propagation of gravitational waves in scalar-tensor theory. Using one simple laboratory of axially symmetric collapse -- the collapse to a black hole of a non-minimally coupled cosmic string loop -- we have extracted the rich spectrum of the scalar-tensor gravitational waveforms $(\tilde{h}^+,~\tilde{h}^\times,~\tilde{h}^s)$ produced in the event. We emphasise that, while we have picked a specific configuration and gravitational theory, the results we have found are general to theories that involve the propagation of a (minimally or non-minimally coupled) massive scalar field. We also expect similar results for theories in which the speed of propagation of the tensor field has a non-trivial  dispersion relation.

    One of the advantages of working with configurations which are not spherically symmetric is that we also produce (massless) tensor gravitational waves that behave in a well-established fashion, and we can compare these to the additionally excited {\it massive}, scalar sector signal. We have focused on the key features by extracting the scalar waveforms at different radii: (i) a faster decay than tensor waves ($r^{-1}$); (ii) a broadening of the wavepacket; and (iii) a delay in the time of arrival for different frequency waves due to mixing of waves emitted during different stages of the event. The fact that the wave packet spreads and the peak amplitude decays more quickly than for the tensor gravitational waves means that massive scalar gravitational waves will be much fainter (see Fig. \ref{fig:dilaton_powerenergy}), and thus, harder to detect. They can carry a comparable amount of energy away from the system as the tensor modes, and this energy is conserved as they propagate, but their luminosity is orders of magnitude fainter than tensor modes due to dispersion.
    
    We have also used this exotic collapse to delve into the generation of both tensor and scalar QNMs from the Schwarzschild black hole. For extracted tensor $\tilde{h}^+$ waveforms, we can clearly identify a dominant QNM at the predicted frequency. The main obstacle to reproduce a similar analysis for the scalar QNM is the dispersive nature of the evolution of the scalar waves themselves, due to mixing with waves emitted during louder stages of the collapse. We have described a simple but powerful method to alleviate this issue, using an effective massive flat space wave equation to \textit{rewind} back the extracted signals to reconstruct their non-dispersed waveforms. This has allowed us to separate out and classify the collapse signal into the \textit{pre-BH formation}, \textit{ringdown} and \textit{post-BH formation} stages. Focusing on the ringdown phase, we have demonstrated excellent agreement with the QNM frequencies expected from perturbative calculations.
    
    Our analysis reinforces what we actually mean by the QNMs around a black hole - the QNM (complex) frequencies are those excited at the perturbed horizon and not those extracted at large distances. In the case of massless tensors (and massless scalars), the shape of the QNM (in the form $e^{t/\tau}\cos(\omega t+\phi)$) is preserved as it propagates away from the black hole. But, as we see in our set up, this is not the case for a massive wave. This means that it is incorrect to assume that at detection, the massive waveforms will have the canonical, QNM shape which is normally considered. Thus some of the assumptions that have gone into forecasting the observability of such QNMs in previous work are incorrect \cite{Tattersall:2019pvx}.
    
    Even though the properties of scalar gravitational waves mean that it is extremely unlikely to detect such a new polarisation mode directly, 
    it is interesting to note that they contain extremely rich information about the theory, the propagation, and the strong gravity event.
    If we (optimistically) assume that we are indeed able to detect the massive scalar waves from a nearby event, then we can imagine rewinding its propagation until the QNM frequencies are consistent with the scalar mass, allowing us to potentially infer, independently, both the scalar mass and the distance to the event. This could be used to break degeneracies in the properties of the collapse process, or the parameters of the final black hole. Furthermore, scalar gravitational waves could be an interesting laboratory in which to test the generation of \textit{non-linearities}, where QNMs interact and additional modes can be sourced \cite{Gleiser:1996yc,Zlochower:2003yh,Bantilan:2012vu,Sberna:2021eui}. It is possible that the generation and propagation of massive scalar waves could result in the excitation of those non-linear mode-couplings.
    
    We end by reiterating that the phenomena we have studied are much more general than the specific scalar-tensor theory, and source, that we studied here. General scalar-tensor theories, such as Horndeski theory \cite{Horndeski:1974wa,Deffayet:2009wt}, its extensions \cite{Gleyzes:2014dya,Zumalacarregui:2013pma} and particular notable examples like Einstein-Dilaton-Gauss-Bonnet gravity \cite{Sotiriou:2013qea} and Chern-Simons gravity \cite{Alexander:2009tp}, all give rise to scalar waves that are potentially massive. In addition, similar effects will be relevant and should be taken into account in theories in which tensor waves have a dispersion relation \cite{deRham:2018red,Baker:2022rhh}. Understanding the rich phenomenology of the excitation and propagation during strong gravity events provides both an opportunity and a challenge, as a route to uncovering distinctive signatures of deviations from general relativity. \\
    
    \section*{Acknowledgments}

    \noindent We acknowledge useful conversations with Jamie Bamber, Amelia Drew, Thomas Helfer, Christopher Moore, Miren Radia and Masahide Yamaguchi. This project has received funding from the European Research Council (ERC) under the European Union’s Horizon 2020 research and innovation programme (grant agreement No 693024). JCA acknowledges funding from the Beecroft Trust and The Queen’s College via an extraordinary
    Junior Research Fellowship (eJRF). PGF acknowledges support from STFC, the Beecroft Trust and the ERC.
    KC acknowledges funding from the ERC, and an STFC Ernest Rutherford Fellowship project reference ST/V003240/1. For the purpose of Open Access, the author has applied a CC BY public copyright licence to any Author Accepted Manuscript version arising from this submission.
    
    The simulations presented in this paper were  performed using PRACE resources under Grant Number 2020225359 on the GCS Supercomputer JUWELS at the Jülich Supercomputing Centre (JCS) through the John von Neumann Institute for Computing (NIC), funded by the Gauss Centre for Supercomputing e.V. (www.gauss-centre.eu), COSMA7 in Durham and  Leicester DiAL3 HPC under DiRAC RAC13 Grant ACTP238.
    
    \bibliography{mybib}
    
    % Remove the % below for adding the appendix
    
    \appendix
    
    \counterwithin{figure}{section}
    
    \section*{Numerical methodology}

    \subsection*{Evolution equations}
    
    In this work, we use \textsc{GRChombo}, a multipurpose numerical relativity code \cite{Clough:2015sqa,Andrade:2021rbd,Radia:2021smk} which solves the BSSN \cite{Baumgarte:1998te,PhysRevD.52.5428} and CCZ4 \cite{Alic:2011gg,Alic:2013xsa} formulations of the Einstein equations. In addition, we use the standard moving puncture gauge conditions \cite{Bona:1994dr,Campanelli:2005dd,Baker:2005vv} for numerically stable evolutions of black hole spacetimes.\\
    
    Having as starting point the action in Eqn. \eqref{eq:act_einst}, the equations of motion for the scalar field are given by Eqn. \eqref{eq:eom}. It is often useful to split the two degrees of freedom of the complex scalar field into a pair $\phi_a = (\phi_1,\phi_2)$ of real scalar fields $\Phi= \phi_1 + i\phi_2$. We write the matter evolution equation as two first order equations with BSSN variables
    \begin{align}
    \partial_t \phi_a =& \alpha \Pi_{a} +\beta^i\partial_i \phi_a ,
    \\
    \partial_t \Pi_{a}=&\beta^i\partial_i \Pi_{a} + \gamma^{ij}(\alpha\partial_i\partial_j \phi_a + \partial_i \phi_a\partial_j \alpha) \nonumber\\
     &+\alpha\left(K\Pi_{a}-\gamma^{ij}\Gamma^k_{ij}\partial_k \phi_a\right)\nonumber\\
     &+\alpha \tilde{\Gamma}^a_{bc}\left(\gamma^{ij}\partial_i\phi_b \partial_j\phi_c - \Pi_b \Pi_c\right)
     + \alpha G^{ab}\frac{\partial \hat{V}}{\partial \phi_b} ,
    \end{align}
    where the field-space Christoffel symbols $\tilde{\Gamma}^a_{bc}$ are defined via Eqn. \eqref{eq:field_chris} with
     \begin{align}
        G_{ab}&=\frac{\mpl^2}{\left(\mpl^2-\frac{\alpha}{6}\left(\phi_1^2+\phi_2^2\right)\right)^2}\times  \nonumber\\ 
        &\begin{pmatrix}
         \mpl^2-\frac{\alpha}{6}(1-\alpha)\phi_1^2 - \frac{\alpha}{6}\phi_2^2 & \frac{\alpha^2}{6}\phi_1\phi_2\\ 
         \frac{\alpha^2}{6}\phi_1\phi_2 & \mpl^2 - \frac{\alpha}{6}\phi_1^2 -\frac{\alpha}{6}(1-\alpha)\phi_2^2 \nonumber
        \end{pmatrix}
    \end{align}

    \subsection*{Initial data}
    
    We find the field theory solution of static infinite string  using a cylindrically symmetric ansatz $\Phi(\mathbf{r})=f(r)\exp(i\theta)$ . We solve the field theory equations numerically for the profiles $f(r)$ \cite{Abrikosov:1956sx,Nielsen:1973cs,Kibble:1976sj,Vilenkin:1981kz,Vilenkin:2000jqa}, with boundary conditions
    \begin{align}\label{eq:bcs}
    f(r\rightarrow 0)=0, \qquad
    f(r\rightarrow \infty)=\eta.
    \end{align}
    We can then construct a field theory loop $\hat{\Phi}$ using a single static string solution via the product rule ansatz \cite{Shellard:1987bv, MULLERHARTMANN1966521}
    \begin{equation}
    \hat{\Phi} = \Phi_1(\mathbf{r}-\mathbf{r_1}) \bar{\Phi}_2(\mathbf{r}-\mathbf{r_2}) ,
    \end{equation}
    where $\Phi_i(\mathbf{r}-\mathbf{r_i})$ is the profile of a single static string with core at $\mathbf{r_i}$, playing the role of the radius $R_0$ of the loop.\\

    The initial data needs to satisfy the Hamiltonian and momentum constraint equations. For simplicity, we choose an initially stationary loop, so that the momentum constraints are trivially satisfied, and we then can solve only the Hamiltonian constraint. We choose a vanishing trace of the extrinsic curvature $K=0$ and a conformally flat ansatz for the $3$-metric
    \begin{equation}
        dl^2 = \psi^4 (dx^2+dy^2+dz^2),
    \end{equation}
    so that the Hamiltonian constraint further reduces to a differential equation for the conformal factor $\psi$,
    \begin{equation} 
    \delta^{ij}\partial_i\partial_j\psi =- 2\pi\psi^5\rho,
    \end{equation}
    sourced by the energy density of the initial configuration
    \begin{align}
        \rho = \frac{G_{ab}}{2}\left(\Pi_a\Pi_b + \gamma^{ij}\partial_i\phi_a\partial_j\phi_b\right) + \hat{V} .
    \end{align}
    
    \begin{figure}[h!]
        \includegraphics[width=\columnwidth]{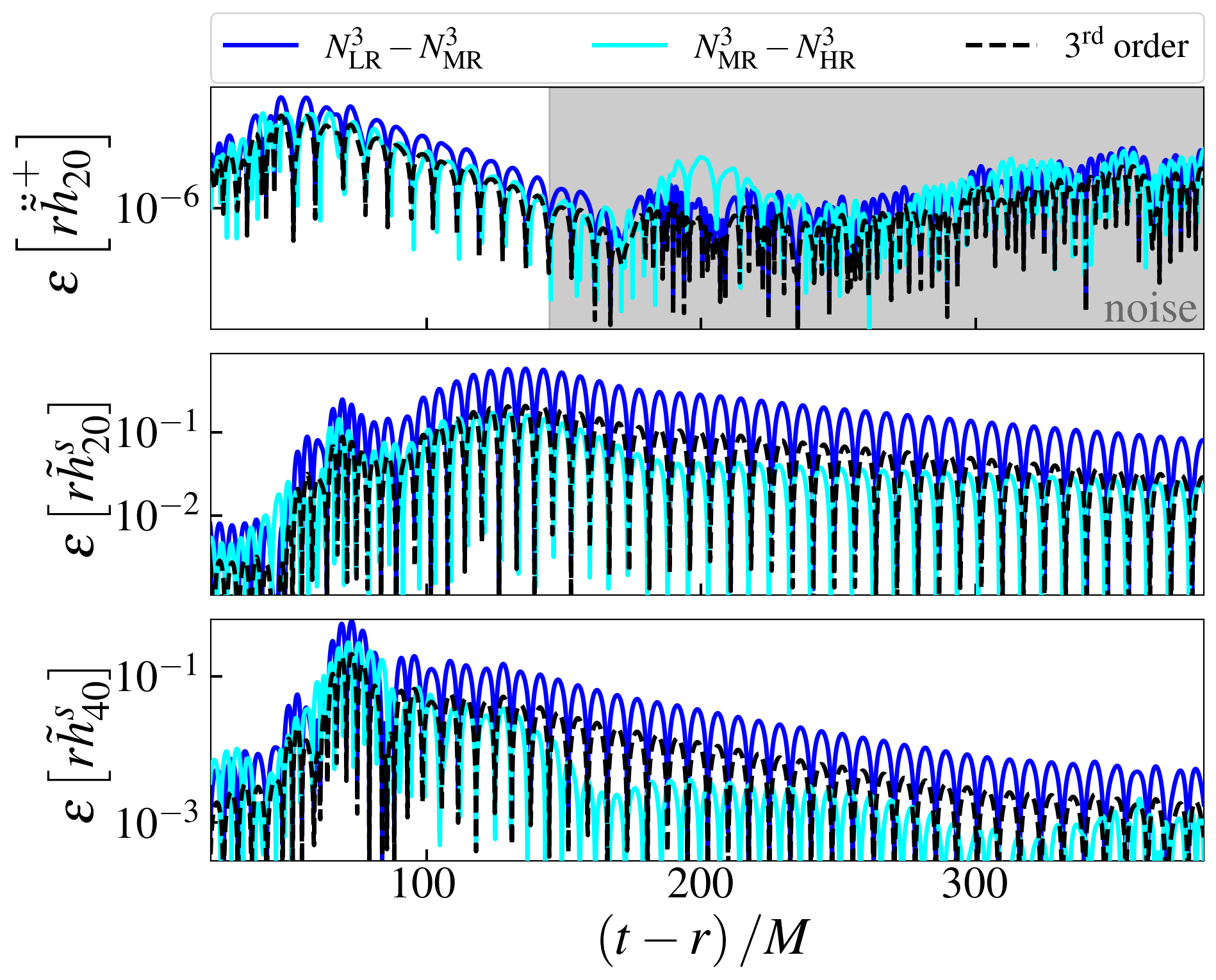}
        \caption{Convergence test for tensor and scalar waveforms using 6 adaptive mesh refinement levels for three coarser resolutions $(N_\mathrm{LR},N_\mathrm{MR},N_\mathrm{HR})=(384,512,640)$. The error in the waveforms $\epsilon$ decreases at a rate consistent with $3^\mathrm{rd}$ order convergence.}
        \label{fig:convergence_test}
    \end{figure}
    
    \subsection*{Convergence testing}
    
    We simulate the system in a box of length $L=12800\approx 545M$ using $6$ refinement levels (2:1 ratio) with reflective boundary conditions, so that we only evolve $1/8$ of the box, considerably speeding up and decreasing the computing resources needed. We convergence test our results by comparing the extracted tensor and scalar waveforms at $\rext=125M$ for three different coarse grid resolutions high $N^3_\mathrm{HR}=640^3$; medium $N^3_\mathrm{HR}=512^3$; and low $N^3_\mathrm{HR}=384^3$. We define an error function which compares pairs of waveforms with different resolutions
    \begin{align}
        \epsilon[f] = \mathrm{abs}(f_\mathrm{N_1} - f_\mathrm{N_2}),
    \end{align}
    which we plot in Fig. \ref{fig:convergence_test}, showing that the error decreases in agreement with $3^\mathrm{rd}$ order convergence. This loss from the used $4^\mathrm{th}$ order stencils is expected due to the AMR regridding.

    \end{document}